# A faint galaxy redshift survey to $B = 24$

Karl Glazebrook,[1] Richard Ellis,[2] Matthew Colless,[3] Tom Broadhurst,[4]
Jeremy Allington-Smith[1] and Nial Tanvir.[2]

[1] *Department of Physics, University of Durham, Science Laboratories, South Road, Durham DH1 3LE**
[2] *Institute of Astronomy, Madingley Road, Cambridge CB3 0HA*
[3] *Mt. Stromlo & Siding Spring Observatories, Australian National University, Weston Creek, A.C.T. 2611, Australia*
[4] *Department of Physics and Astronomy, The John Hopkins University, Baltimore, MD 21218, USA*

**ABSTRACT**
Using the multislit LDSS-2 spectrograph on the *William Herschel Telescope* we have completed a redshift survey in the magnitude range $22.5 < B < 24$ which has produced 73 redshifts representing a 73% complete sample uniformly-selected from four deep fields at high Galactic latitude. The survey extends out to $z > 1$ and includes the highest redshift galaxy ($z = 1.108$) yet discovered in a field sample. The median redshift, $z_{MED} = 0.46$, and form of the redshift distribution constitute compelling evidence against simple luminosity evolution as an explanation of the large excess of faint galaxies ($\simeq \times 2$–4 no-evolution) seen in this magnitude range. Rather we identify the excess population as blue objects with $z \sim 0.4$ and $B$ luminosities similar to local $L^*$ galaxies indicating a dramatic decrease in the density of such objects over the last Hubble time, confirming the trends found in brighter redshift surveys. We also find a marked absence of *very* low redshift galaxies ($z < 0.1$) at faint limits, severely constraining any significant steepening of the local field galaxy luminosity function at low luminosities.

**Key words:**

## 1 INTRODUCTION

The number-magnitude and number-redshift distributions of faint ($B > 20$) galaxies are important probes of the geometry and evolution of the universe. It is now well-established that the number counts increasingly exceed the prediction from a non-evolving galaxy population fainter than $B \sim 20$ (Jones *et al.* 1991, Metcalfe *et al.* 1991, Lilly, Cowie & Gardner 1991, Tyson 1988). One possible explanation for this excess is that star-formation in galaxies was higher in the past and that consequently galaxies were more luminous. Since they would be seen to higher redshifts than predicted from a 'no-evolution' model, a greater projected space density would result. A more radical possibility is that the comoving space density of visible galaxies has not been conserved, either because a dwarf population visible at earlier epochs has now faded beyond detection (Cowie 1991, Cowie, Songalia & Hu 1991, Babul & Rees 1992), or because present-day galaxies have been formed through self-similar merging of more numerous fragments (Rocca-Volmerange & Guiderdoni 1990, Broadhurst, Ellis & Glazebrook 1992, hereafter BEG).

To match the high surface density of galaxies found at faint magnitudes, luminosity evolution models inevitably predict a high mean redshift for the galaxies, with a substantial fraction beyond $z \simeq 1$ at $B = 24$. However, redshift surveys of several hundred galaxies to $B = 22.5$ (Broadhurst, Ellis & Shanks 1988, Colless *et al.* 1990, 1993) appear to show a *deficit* of high-redshift galaxies compared to the predictions of such models. Recently some fields have been surveyed to very high completeness limits, eliminating the possibility that a significant fraction of unidentified sources lay beyond $z \simeq 1$ (Colless *et al.* 1993). Simple luminosity evolution models also predict that the very blue galaxies found in significant numbers fainter than $B \approx 22$ should almost all be at redshifts $z > 1$ (c.f. Koo & Kron 1992), whereas the Colless *et al.* study of a small sampled of 'flat-spectrum' galaxies found that all had $z < 1$.

These surveys constrain *luminosity* evolution at the bright end of the luminosity function and point to some form of evolution in *number density*. As such, the interpretation remains controversial. Metcalfe *et al.* (1991) and Koo & Kron (1992) have argued that, to $B = 22.5$, it is still possible to reconcile the observed redshift distributions with density-conserving mild luminosity evolution models, especially if the bright counts for $B < 19$ (Maddox *et al.* 1990, Heydon-

---

* Present address: Institute of Astronomy, Madingley Road, Cambridge CB3 0HA

Dumbleton, Collins & MacGillivray 1989) are neglected and a high density normalization for the no-evolution model is used.

More recently Koo, Gronwall & Bruzual (1993) have introduced a no-evolution model empirically 'tuned' to fit the faint counts and redshift distributions by invoking a steep local luminosity function for dwarf blue galaxies. This model has been proposed before, the various flavours are discussed extensively in Broadhurst *et al.* (1988).

Since the divergence between the various types of model rapidly increases at fainter apparent magnitudes, a critical test is the redshift distribution of a large sample of galaxies at a yet fainter limit. The only published work fainter than $B = 22.5$ is that of Cowie *et al.* (1991) who measured the redshift of 12 galaxies in this range — here we report the results of a much larger redshift survey to $B = 24$ carried out using the new LDSS-2 multislit spectrograph on the William Herschel Telescope on La Palma. Section 2 describes the selection of the sample and the spectroscopic observations. The results of the survey are given in Section 3 and discussed in terms of their implications for models of galaxy evolution in Section 4. Our conclusions are summarized in Section 5.

## 2  OBSERVATIONS

### 2.1  Photometric observations

The galaxies were selected from deep $B$ and $R$ CCD images of four equatorial fields. The details of the observations including coordinates and field areas are listed in Table 1. The images were obtained using the TAURUS f/4 focal reducer and EEV large format CCD on the William Herschel Telescope in May 1991 and on the Isaac Newton Telescope using the RCA CCD in September 1989.

Standard debiasing and sky flatfielding reductions were applied. The TAURUS focal reducer suffers from a radial astrometric distortion of up to $3''$ across the field (Smail 1993) which was corrected for each individual frame by resampling the data to a linear grid before coaddition. On each frame, 20–30 objects were selected for which astrometric positions were known from wide-field AAT prime focus plates. Using these coordinates, a 2-D spline was fitted to the distortion pattern leaving residuals $< 0.2''$. The frames were then resampled removing the non-linear term. Because the distortion is small compared to the size of the field, the photometric correction arising from this procedure is small ($\Delta m \approx 0.01$). Its main importance is in the need for precise astrometry of target objects for the LDSS-2 multislit masks.

Objects were found automatically by first smoothing the frames with a Gaussian of width equal to the seeing, and running the FOCAS software with an area threshold equal to the seeing disk and a flux threshold $3\sigma$ ($\equiv 26.7\,\text{mags/arcsec}^2$) above the background. Aperture photometry was performed in a $4''$ diameter aperture, to allow for the seeing an aperture correction of $\sim -0.3$ mags (exact value determined independently in each field from bright stars) was applied. Obviously for extended objects there will be a further correction, this is explored further in Section 4.

As some of the data was taken in non-photometric conditions, it was calibrated with reference to the brighter CCD data of Glazebrook *et al.* (1994). Our initial estimates based on the faint isophotes used for image detection were that the detections should be complete to $B > 24$ — comparison with the published data (Metcalfe *et al.* 1991) show our mean counts to be in excellent agreement to $B = 24$ where the random photometric errors are $\sim 0.1$ mag. For $22.5 < B < 24$ we count 17 200 galaxies/deg$^2$ with a field–field RMS variation of 40%. This variation is slightly higher than expected — following the method of Glazebrook *et al.* (1994) we estimate from galaxy clustering that independent areas of size 50 arcmin$^2$ should have 25% RMS fluctuations. However with only 4 small fields we do not attribute significance to this discrepancy.

In selecting the spectroscopic sample a bright cut was made at $B = 22.5$ so that the entire sample would lie beyond the faint limit of the earlier LDSS-1 surveys (Colless *et al.* 1990, 1993). This resulted in a B-selected catalogue over the four fields of 1002 objects in the range $22.5 < B < 24$. As in the LDSS-1 surveys, *the object selection was based purely on apparent magnitude* — no star-galaxy separation was attempted. Since we expect *a priori* relatively few stars at these faint magnitudes this will hardly affect the efficiency in measuring galaxy redshifts but guards against exclusion of compact extragalactic sources. Our results also allows us to measure the number of high redshift QSOs at these magnitudes. The completeness limits and numbers of objects in the individual fields within the selected magnitude range ($22.5 < B < 24$) are given in Table 1.

### 2.2  Spectroscopic observations

The LDSS-2 faint object spectrograph is very similar to the LDSS-1 instrument described by Wynne & Worswick (1988) and Colless *et al.* (1990). A full description is given by Allington-Smith *et al.* (1994). Briefly, LDSS-2 can accept multislit masks with slits positioned anywhere over a field of diameter $11.5'$. A choice of dispersions is available and there is an imaging mode, primarily for the purposes of field acquisition. The main improvements over LDSS-1 are automation of all functions, better optics with an improved point-spread function and more filter and grism options.

For the observations reported here, seven masks were used: three in the $10^{\rm h}$ field, two in the $13^{\rm h}$ field and one in each of the other fields. Each mask had slits cut for between 20 and 34 objects. The slits are of length $15$–$20''$, and arranged so that the dispersed images do not overlap in either the spectral or spatial directions.

LDSS-2 was commissioned in March–April 1992 and during that period three masks were observed as part of the science verification. Two more masks were observed during a second run in August–September 1992 and one of the first three re-observed in March 1993. A further set of masks were observed in April 1994 to extend the sample, with objects chosen in areas of forthcoming *Hubble Space Telescope* observations. A brighter lower limit ($B > 20$) was used for these masks. For the rest of this paper we refer only to the $22.5 < B < 24$ subsample with these masks, although all of the identifications are later given in Table 3.

For all three runs a dispersion of $5.3$Å per pixel was used with the grating blazed at $5000$Å. The detector was a Tektronix $1024 \times 1024$ CCD with a peak quantum efficiency of 85% in the red, dropping to 40% at $4000$Å. The peak measured throughput of LDSS-2 including this CCD is $\simeq 22\%$ at 6000 Å, which, combined with the chip's readout noise of 4 $e^-$, meant that exposures had to be at least 1800 s long in order to be sky-limited. The spectroscopic observations are summarized in Table 2. Both the throughput and spectral resolution are a considerable improvements over that of LDSS-1 used in the earlier AAT surveys. The atmospheric seeing was also generally better than that during the AAT runs. Together with the improved optics, these factors have made it relatively straightforward to push the limiting magnitude from $B=22.5$ to $B=24$.

The individual 1800 s observations were combined with a cosmic ray filter to give final summed images. The spectra were optimally extracted according to the seeing in individual frames and calibrated using the LEXT software described in Colless *et al.* (1990). Both the 1-D and 2-D spectral information was used in confirming the reality of emission lines. Each of the spectral identifications were confirmed via independent examinations by four of the authors (KGB, RSE, MMC and TJB).

## 3 RESULTS

### 3.1 The redshift survey

In our $22.5 < B < 24$ sample we have observed a total of 157 objects, and identified 84 galaxies (with redshifts $0.081 \leq z \leq 1.108$), 8 stars and 2 QSOs. Additionally brighter than that $B = 22.5$, we have found 11 new galaxy redshifts.

For each object in the survey, Table 3 lists the mask, slit number, unique object ID number, sky coordinates, optical magnitudes and, where obtained, the redshift and rest-frame equivalent width (or upper-limit thereof) of [OII] 3727Å. Also given is a 'quality' criterion similar to that originally defined by Colless *et al.* (1990). $Q = 1$ indicates a reliable identification based on more than one feature, $Q = 2$ an identification based on a single feature (usually [OII] 3727Å or Ca H+K), $Q = 3$ indicates those spectra for which no identification was possible and $Q = 4$ those for which no spectrum at all was detected. A spectral type of 'E', 'A', or 'EA' is given depending on whether emission or absorption features or both are seen. If the object is a suspected QSO the type is given as 'Q'. Stars are given as 'S'. The comments column of Table 3 lists the features found in each spectrum.

To check whether the $Q = 4$ objects were genuine and not artifacts in our original CCD data we obtained a repeat image of the $13^h$ field in April 1994. All the objects for which we attempted to obtain redshifts, including the $Q = 4$ objects were seen again. All the $Q = 4$ objects lie close to the magnitude limit of the survey, so we conclude that the spectra were not detected because they are just the extreme examples of $Q = 3$ objects with weak continua. Thus we include these objects in our calculations of the incompleteness, which is indeed the conservative assumption. We note that leaving them out would raise the overall completeness to $\simeq 80\%$ and strengthen further the conclusions presented later.

### 3.2 Completeness and reliability of spectroscopic identifications

It can be seen from Table 2 that four of the masks are $\gtrsim 70\%$ complete to $B = 24$, while the remaining three are less complete. This is primarily because the latter masks were observed in rather poorer conditions of seeing and/or transparency leading to inadequate $S/N$ for faint objects. To allow for this we estimate from the data the limiting magnitude $B_{70\%}$ at which the completeness is $\gtrsim 70\%$. We then take this as the appropriate magnitude limit for these two fields — it is given in Table 2. We note that fainter than $B_{70\%}$ the identified and unidentified objects have the same $B - R$ colour distribution.

In our final $B < B_{70\%}$ sample there are 111 objects in total, 81 of which have identifications so the overall completeness is 73%. Of these 73 are galaxies, 6 are faint stars and 2 we identify as QSOs owing to their high redshift and broad lines. For our later analysis we also consider a smaller but more complete sub-sample: if we use only the 4 best masks (03z3_A and the $10^h$ masks) and take a uniform magnitude cut of $22.5 < B \leq 23.5$ then we have a 89% complete sub-sample (30 galaxies, 2 QSOs, 1 star and 4 unidentified). We examine these samples in Section 4.

62 objects have $Q = 1$ and 19 have $Q = 2$. The mask 10z2_B was observed twice, the second time in much better conditions with a much greater limiting depth. This gives us a valuable internal check on the reliability of our quality values. The second observation was reduced a year after the first with no reference to the earlier notes until after the identifications were established. The results of this exercise are illuminating — all 4 of the original $Q = 1$ identifications and 4 out of 6 of the $Q = 2$ identifications on the earlier observation were correct. All the [OII] equivalent widths were the same within the errors. This gives us confidence in the overall reliability of our identifications.

As it is impractical to show all 81 spectra here we have chosen 11 of them, using a random number generator, to demonstrate the typical quality of the spectra and the reliability of the identifications. The spectra (shown in Figure 1) are optimally smoothed with a 15 Å FWHM Gaussian to match the instrumental response. The sky-subtraction is occasionally poor near the bright sky lines at 5577Å, 5892Å and 6300Å so these parts of the spectrum, together with the odd residual cosmic ray, have been blanked out in the figure. Our line identifications are marked. It should be re-emphasized that our identifications do not rest solely on the 1D spectra — the 2D sky-subtracted images were also scrutinized to distinguish bright emission lines from cosmic rays.

The remaining twelfth object in Figure 1 (top left, 10.288) is not randomly chosen — we show it because it is the highest redshift galaxy yet seen in a published field sample ($z = 1.108$). It shows strong extended [OII] emission and, interestingly, several absorption features which are clearly identified with MgII, MgI and FeII. It is a much more secure identification than the $z = 1.018$ object of Thompson & Djorgovski (1991) whose spectrum only revealed [OII] and was, in any case, close to a QSO. The MgII/MgI equivalent width ratio is typical of those seen in the stronger Mg absorption line systems (Steidel & Sargent 1992) and the $B - R = 0.3$ colour indicates that this is a flat spectrum object. Neither its luminosity, $M_B \simeq -18.8 \simeq 0.4 L^*$ assum-ing flat spectrum ($\Delta f_\nu = 0$) K-corrections, nor its redshift is particularly unusual. As we will show in Section 4, the object simply represents the high-redshift tail of the distribution of our deep data, even without any evolution in luminosity.

## 4 DISCUSSION

The magnitude-redshift relation for all the identified galaxies is shown in Figure 2, where we have also plotted previously published B-selected field samples (Peterson *et al.* 1986, Broadhurst *et al.* 1988, Ellis & Broadhurst 1993, Colless *et al.* 1990 & Cowie *et al.* 1991). The deepest previous data is provided by Cowie *et al.* (1991) who measured 12 redshifts forming a complete sample in the range $22.5 \leq B \leq 24$ acquired via single-slit spectroscopy with an average integration time of 15 000 s per object. Other deep surveys have been published recently but these involve selection in bands other than $B$. Colless *et al.* (1993) obtained 11 redshifts for a sample of flat spectrum galaxies with $22 < R < 23$ with LDSS-1, and Lilly (1993) and Tresse *et al.* (1993) have various $I$-limited surveys underway. Concentrating on the $B$-limited samples, the figure shows a smooth trend, both in terms of the upper redshift envelope and the mean redshift increasing slowly for progressively fainter limits.

### 4.1 Faint QSO number-densities

Before discussing the detailed properties of the galaxy population we consider the number of QSOs found. The observed fraction of QSOs corresponds to a number density of $266^{+572}_{-213}$ deg$^{-2}$ mag$^{-1}$ (95% confidence limits calculated using Poisson statistics as in Gehrels 1986) at $B \simeq 23.2$ (mean magnitude of sample). The lower limit is, of course, more secure than the upper limit due to the number of unidentified objects in the sample, though QSOs with their strong emission lines (e.g. MgII, CIII], CIV) in the UV should be easier to identify than galaxies.

QSO number-magnitude counts at bright magnitudes ($B < 20$) (summarized by Hartwick & Schade 1990) are well fit by a power-law with slope 0.86. Extrapolated to $B = 23.2$ this would predict 26 000 deg$^{-2}$ mag$^{-1}$ which is far in excess of our limits, even allowing for the uncertainties. In contrast our new survey, as that of Colless *et al.* (1991), supports the turnover in QSO counts defined by Boyle, Shanks & Peterson (1988) and Koo, Kron & Cudworth (1986) at $B = 20$. While our formal errors are somewhat larger than those of the compact objects survey of Colless *et al.*, due to their larger sample, our independent result confirms their measurement in a fainter sample with no compactness criterion. Extrapolation of the flatter slope beyond $B=20$ predicts only $80\,\mathrm{deg}^{-2}$ mag$^{-1}$ at $B = 23.2$. While consistent with our numbers, there is the first indication of a somewhat larger number of QSOs than that found from UVX techniques alone (c.f. Hawking & Veron 1993).

### 4.2 Faint galaxy colours

Of the 111 objects in the $B < B_{70\%}$ sample, 108 have $R$ data and 100 have reliable $B - R$ colours to a limit of $R = 24$. Figure 3 shows the colour-redshift relation. On the left of the main plot we plot the colours of the stars, and on the right we plot those of the unidentified objects (though we do not mean to suggest that they are high-redshift objects). We also plot no-evolution loci for standard Hubble types, although we do see early-type galaxies it is immediately obvious that the population extends considerably bluer than the templates and many have flat-spectrum colours ($B - R = 0.2$). The number of galaxies in the $22.5 < B < 24$ sample is a factor of 2–4 × the no evolution prediction and the bulk of this blue excess is clearly identified as galaxies with $0.2 < z < 0.7$.

It is also clear that the unidentified objects in the $B < B_{70\%}$ sample are on average bluer than the identified sample and occupy a relatively narrow range in colour. For the purpose of constraining any evolutionary tail, it is important to determine the true redshift distribution of the unidentified population. We consider three possibilities:

(i) They could be very low-redshift ($z < 0.2$) low-luminosity systems. Although they have the appropriate colours, such objects would normally show spectra typical of HII regions with strong emission lines of [OII], H$\beta$, [OIII] & H$\alpha$. In our spectral window these should be the *easiest* objects to identify — not the hardest. The only possibility would be if they represented a class of object whose star formation had just ceased. However in that case we would also expect to see many red systems with $z < 0.2$, so we conclude this explanation is most unlikely.

(ii) They could have the same redshift distribution as the data ($0 < z < 1$). In this respect, it is curious to know why their redshifts were not determined. Figure 4 demonstrates a strong correlation between the [OII] equivalent width and $B - R$ colour in our sample. This may not be a selection effect arising from the absence of identified weak [OII] systems since low-redshift samples (e.g. Kennicut 1992) also show the correlation. However we cannot firmly rule out this possibility and note that Colless *et al.* (1993) showed that the bluest systems in the unidentified LDSS1 sample had $0 < z < 1$.

(iii) The final possibility is that the unidentified objects are at high-redshifts, $z > 1$, and the reason they remain unidentified is because [OII] is redshifted out of the spectral window leaving only weak absorption features (MgII, MgI) which are difficult to identify at low $S/N$. Considering the earlier work of Colless *et al.* (1993), we do not consider this likely as the sole explanation since they found no such examples.

None of these hypotheses is convincing as a sole explanation, we expect that the most likely answer is a combination of (ii) and (iii) as Colless *et al.* (1993) found some blue weak-lined [OII] systems at $z < 1$ and even in a no-evolution model we expect a number of galaxies at $B = 24$ to lie at $z > 1$ (see Section 4.3). When testing against models of luminosity evolution, as we will do in Section 4.3, the most conservative assumption is to place all the unidentified galaxies at $z > 1$ which gives an upper limit on the proportion of these galaxies. As we will demonstrate our sample is sufficiently large and complete that our conclusions are not significantly altered by the placement of these galaxies.

### 4.3 The redshift distribution

The redshift distribution of the 73 galaxies in our 73% complete sub-sample is shown in Figure 5 together with the pre-

### 4.3.1 The adopted zero redshift luminosity function

We use a more recent luminosity function than BEG with the Schechter parameters $(M^*, \alpha)$ for early and late type galaxies taken from Loveday *et al.* (1992) and a morphological mix adjusted to match the distribution of types seen at $b_J < 16.7$ by Shanks *et al.* (1984). We note that it has recently been suggested by Zucca, Pozzetti & Zamorant (1994) that the Loveday *et al.* analysis is in error and on reanalysis Zucca *et al.* get Schechter parameters closer to the older values of Efstathiou, Ellis & Peterson (1988). Our analysis below has been duplicated using the Efstathiou *et al.* $(M^*, \alpha)$ parameters: the conclusions remain unchanged.

The absolute normalisation of the zero redshift luminosity function, $\phi^*$ currently remains uncertain by a factor of two. If this is set higher then this is equivalent to normalising the no-evolution number-magnitude curve at a fainter apparent magnitude; and lowers the excess of faint blue galaxies to be explained. Loveday *et al.* find a value of $\phi^* = 0.015\, h^3\, \mathrm{Mpc}^{-3}$ from their luminosity function analysis based on bright $b_j < 17$ APM data. Metcalfe *et al.* (1991) argued that the normalisation should be at $b_J = 19$ (equivalent to $\phi^* = 0.03\, h^3\, \mathrm{Mpc}^{-3}$) as the brighter data could be subject to local density fluctuations or calibration effects. These two values bracket the range of estimates in the literature; to be conservative we adopt the higher value.

The model predictions are calculated allowing for the variation of $B_{70\%}$ between fields. To normalise the models we compute an effective area for each field based on its magnitude limits and assuming a random sampling of the known number-magnitude counts. To allow for our field-field number fluctuations we normalise to the number-magnitude counts of Metcalfe *et al.* (1991), Jones *et al.* (1991) and Lilly *et al.* (1991) which are much better determined over a larger area than in our survey, though our mean $22.5 < b_j < 24$ counts agrees well. Over the narrow range $22 < b_J < 25$ we find $\log(N/\mathrm{mag}^{-1}\,\mathrm{deg}^{-2}) = 2.62 + 0.43(b_J - 20)$ to be an excellent empirical fit to these data. Values for the effective areas thus calculated are given for each field in Table 2. For the more complete sub-sample the total effective area is 14.2 arcmin$^2$.

### 4.3.2 Correction for aperture effects

One further issue is the aperture correction, *i.e.* the correction of the flux from the measured aperture to some notional "total". The conventional approach (*e.g.* Lilly 1993) is to use a small fixed aperture of 3–6$''$ diameter and then correct to total by a fixed offset. Alternatively a fixed isophote is used. Both of these will give redshift-dependent aperture effects, for example our 4$''$ aperture is 8 $h^{-1}$ kpc at $z = 0.2$ and 17 $h^{-1}$ kpc at $z = 1$ for $\Omega = 1$. A fixed isophote, in the observer's frame, gives still more severe effects due to the $(1 + z)^4$ surface brightness dependence.

It would be possible to remeasure our magnitudes in metric apertures now the redshifts are known, but this would destroy the cleaness of our initial $22.5 < B < 24$ selection. Instead we choose to correct our *models* to 4$''$ apertures; this has the additional advantage that all the cosmological dependence is kept in the model. Initially for simplicity we used the growth law of Glazebrook *et al.* (1994b), $L(< r) \propto r^{0.4}$, which is a good approximation to both standard exponential and de Vaucouleurs profiles outside the central few kpc. Indeed Glazebrook *et al.* find this to be an excellent fit to their data. We correct to a standard aperture of 20 $h^{-1}$ kpc which gives a correction ranging from $-0.37$ mags at $z = 0.2$ to $-0.07$ mags at $z = 1$. In practice the effect on the calculated redshift distribution for the range of models considered here turns out to be quite small, typically $\lesssim 0.5$ galaxies per redshift bin at the peak, because the correction is largest at low-redshift where the volume is small. Thus a still more detailed treatment, such as using type-dependent exponential and de Vaucouleurs profiles according to the luminosity function weights, is unnecessary. The statistical results given below are unchanged by use of these aperture corrections.

### 4.3.3 The evolutionary models

BEG define the amount of luminosity evolution via the parameter $b$, which represents different exponential starformation time-scales, normalized so that the low-redshift evolution is $L \propto (1 + bz)$. Curves for $b = 0$ (no-evolution), $b = 2$ and $b = 4$ are plotted. To match the steep slope of the number counts a $b = 4$ luminosity evolution model was preferred by BEG. We also plot a BEG merger model ($b = 3$, $Q = 4$) which best fits the $n(z)$ where the $Q$ parameter defines the rate of increase with look-back time in the number density. This model prediction is not too dissimilar to the no evolution case when renormalized. Finally, we consider the recent model proposed by Koo *et al.* (1993) transformed from $dn/d\log z$ to $dn/dz$. (Note: strictly Koo *et al.*'s model is for $23 < B < 24$ but as this makes negligible difference to the data $n(z)$ we refrain from introducing another figure). Koo *et al.*'s normalisation is not specified in their paper so we choose to scale their curve to match the total number of galaxies and unidentified objects in our sample.

The models are tested against the data in various ways and the results are summarized in Table 4. Firstly we compare the overall shape of the distributions by means of a K-S test ($P_{KS}$). Secondly we consider two less sensitive but more robust statistical tests, which allow us to make statistical statements *despite* the unidentified objects. The first of these considers the distribution of *median* redshifts. If we ignore the unidentified objects, the median redshift of our sample is $z_{MED} = 0.46$. If we assume that all unidentified objects have $z > 1$ then we obtain an upper limit to the median of $z_{MED} = 0.56$. (Similarly if they are all at $z \sim 0$ then $z_{MED} = 0.36$). Clearly we can calculate a median redshift for the model distributions but we need to assign a statistical significance to this. To do this we generated $10^6$ realizations of 100 redshifts (galaxies and unidentified objects) drawn randomly from the model distribution and calculate the probability of observing $0.46 < z_{MED} < 0.56$, which we call $P_{MED}$.

Another interesting question is how does the fraction of high-redshift objects compare between the data and the various models? To quantify this we measure $f_{0.7}$ the fraction of galaxies which have $z > 0.7$. For the data this is 0.12, rising to 0.38 if again we assume that all unidentified objects are at $z > 1$. We calculate $f_{0.7}$ for the models and the prob-ability that $0.12 < f_{0.7} < 0.38$ ($P_{0.7}$) using the realizations which are tabulated in Table 4. Both $P_{MED}$ and $P_{0.7}$ are sensitive to the global distribution over the whole $0 < z < 1$ range and are thus insensitive to galaxy clustering on small scales. For the 89% complete subsample we find a median redshift of 0.46 with limits $(0.41, 0.48)$ and $f_{0.7} = 0.13$ with limits $(0.12, 0.24)$.

It can be seen from Figure 5 that the $b = 2$ and $b = 4$ luminosity evolution models all predict too many high-redshift galaxies, this is confirmed statistically by the tests in Table 4. Note the $b = 2$ model is almost exactly equivalent to the mild luminosity evolution models advocated by Metcalfe et al. (1991) and Koo & Kron (1992) which the authors argue were marginally consistent with the redshift distribution of Colless et al. (1990) given the incompleteness. Colless et al. (1993) reduced the incompleteness and found no $z > 1$ galaxies, moreover the bluest objects were at low redshift. Our extension to $B = 24$ confirms this result and reveals few $z > 1$ galaxies.

Importantly, the data shows a large excess of galaxies at $z \sim 0.4$ with respect to the no-evolution and luminosity evolution models. This is unaffected by the placement of the unidentified objects. Since we have used the highest possible local normalisation of $\phi^*$ this can only be an evolutionary effect. We are clearly seeing an increase in the space density of galaxies with $L \sim L_B^*(z = 0)$. It is impossible for these simple luminosity evolution models to reproduce this as they are only capable of adding extra galaxies above $z > 0.7$.

In contrast the merger model ($b = 3$, $Q = 4$) succeeds in matching the data both in shape and normalisation, primarily because it involves little change in the bright end of the luminosity function while increasing the overall space density. No-evolution models which increase the number of galaxies at the faint end of the local luminosity function, such as that of Koo et al. examined here, produce a marked excess of $z < 0.2$ galaxies *not* seen in our data and do not match the excess of galaxies we *do* see at $z \sim 0.4$. The absence of low-redshift galaxies was already evident in Koo et al.'s own comparison but was rather obscured by their use of a $N - \log z$ plot. This is one of the more significant conclusions arising from the new survey. The paucity of $z < 0.2$ galaxies to $B = 24$ severely constrains any possibility that the faint end of the local galaxy luminosity function has been serverely underestimated as Koo et al. conjectured.

## 5   CONCLUSIONS

The new redshift data presented here are a compelling piece of evidence ruling out simple luminosity evolution as the sole cause of the excess seen in the faint counts. Not only does the lack of high-redshift galaxies in a $B < 24$ sample provide an even more severe limit than in brighter surveys, but the number of galaxies at *low* redshift ($z \sim 0.4$) clearly exceeds the predictions of luminosity evolution models for any reasonable value of the local normalisation. Simple models of luminosity evolution can not rectify this — we rule out any evolution in the bright end of the luminosity function. Importantly, there is also little scope for hypotheses which attempt to explain the faint count excess by modifying the zero redshift luminosity function at the faint end such as proposed by Koo et al. (1993) — LDSS-2 data has identified a *lower* envelope in the $B - z$ diagram which indicates there is no such population. Moreover, such models can not produce the *observed* excess seen at $z \sim 0.4$. Rather, evolution in the space density of blue galaxies with $L \sim L_B^*(z = 0)$ must occur, whatever the underlying mechanism. This could be explained by direct density evolution of galaxies of all luminosities or differential luminosity evolution of only the lower luminosity galaxies up to $L \sim L_B^*(z = 0)$. Thus the conclusion that the total space density of luminous blue galaxies has changed since $z \sim 0.5$–1 becomes unavoidable.

## ACKNOWLEDGMENTS

We would especially like to thank the LDSS-2 team (Mike Breare, Keith Taylor, Dave Gellatly, Graham Shaw, John Webster and Sue Worswick) for the construction of an excellent instrument and for providing support during the commissioning period. Further encouragement from Alec Boksenberg is gratefully acknowledged. The observations were carried out at the *William Herschel Telescope*, operated by the Royal Greenwich Observatory in the Spanish Observatorio del Roque de Los Muchachos of the Instituto de Astrofísica de Canarias. We thank Ian Smail for supplying the software for correcting the TAURUS images, and John Peacock, Chris Collins and Lance Miller for allowing the use of their photometric data. The data reduction was carried out using STARLINK which is funded by the PPARC. KGB, RSE and NRT acknowledge financial support by the PPARC.

## REFERENCES

Allington-Smith J. R., Breare J. M., Ellis R. S., Gellatly D. W., Glazebrook K., Jorden P. R., MacLean J. F., Oates A. P., Shaw G. D., Tanvir N. R. Taylor, K., Taylor P. B., Webster J., Worswick S. P., 1994, PASP, in press
Babul A., Rees M. J., 1992, MNRAS, 255, 346
Boyle B. J., Shanks T., Peterson B. A., 1988, MNRAS, 235, 935
Broadhurst T. J., Ellis R. S., Shanks T., 1988, MNRAS, 235, 827
Broadhurst T. J., Ellis R. S., Glazebrook K., 1992, Nat, 355, 55 (BEG)
Colless M. M., Ellis R. S., Taylor K., Hook R. N., 1990, MNRAS, 244, 408
Colless M. M., Ellis R. S., Taylor K., Shaw G., 1991, MNRAS, 253, 686
Colless M. M., Ellis R. S., Broadhurst T. J., Taylor K., Peterson B. A., 1993, MNRAS, 261, 19
Cowie L. L., 1991, in Shanks T., Banday A. J., Ellis R. S., Frenk C. S., Wolfendale A. W., eds, Observational Tests of Cosmological Inflation, Kluwer, Dordrecht, p257
Cowie L. L, Songalia A., Hu E. M., 1991, Nature, 354, 460
Efstathiou G., Ellis R. S., Peterson, B. A., 1988, MNRAS, 232, 431
Ellis R. S., Broadhurst T. J., 1993, in preparation
Gehrels N., 1986, ApJ, 303, 336
Glazebrook K., Peacock J. A., Collins C. A., Miller L., 1994, MNRAS, 266, 65
Glazebrook K., Peacock J. A., Miller L., Collins C. A., 1994b, MNRAS, in preparation
Hartwick F. D. A., Schade D., 1990, Ann. Rev. Astr. Astrophys., 28, 437
Hawkins M. R. S., Véron P., 1993, MNRAS, 260, 202

Heydon-Dumbleton N. H., Collins C. A., MacGillivray H. T., 1989, MNRAS, 238, 379
Jones L. R., Fong R., Shanks T., Ellis R. S., Peterson, B. A., 1991, MNRAS, 249, 481
Koo D. C., Kron R. G., Cudworth, K. M., 1986, PASP, 98, 397
Koo D. C., Kron R. G., 1992, Ann. Rev. Astr. Astrophys., 30, 613
Koo D. C., Gronwall C., Bruzual G. A., 1993, ApJ, 415, L21
Kennicut R. C., 1992, ApJ, 388, 310
Lilly S. J., Cowie L. L., Gardner J. P., 1991, ApJ, 369, 79
Lilly, S. J., 1993, ApJ, 411, 501
Loveday J., Peterson B. A., Efstathiou G., Maddox S. J., 1992. ApJ, 390, 338
Maddox S. J., Sutherland W. J., Efstathiou G., Loveday J., Peterson B. A., 1990 MNRAS, 247, 1P
Metcalfe N., Shanks T., Fong R., Jones L. R., 1991, MNRAS, 249, 498
Peterson B. A., Ellis R. S., Shanks T., Bean A. J., Fong R., Efstathiou G., Zou Z-L., 1986, MNRAS, 221, 233
Rocca-Volmerange B., Guiderdoni B., 1990, MNRAS, 247, 166
Shanks T. S., Stevenson P. R., Fong R., MacGillivray H. T., 1984, MNRAS, 206, 767
Smail I., 1993, Ph.D. thesis, University of Durham, England
Steidel C. C., Sargent W. L. W., 1992, ApJS, 80, 1
Thompson D. J., Djorgovski S., 1991, ApJ, 371, L55
Tresse L., Hammer F., Le Fèvre O., Proust D., 1993. A&A, submitted
Tyson J. A., 1988, AJ, 96, 1
Wynn C. G., Worswick S. P., 1988, Observatory, 108, 161
Zucca E., Pozzetti L., Zamorant G., 1994, MNRAS, 269, 953

**FIGURE CAPTIONS**

**Figure 1.** Randomly-selected spectra (except for object 10.288) from the LDSS-2 redshift survey showing the claimed features by which they were identified. Gaps in the spectra represent regions where poorly subtracted night sky lines, or occasional CCD defects and residual cosmic rays, have been removed. The spectra have been approximately relatively flux-calibrated (in $f_\lambda$) by dividing by the telescope + instrument throughput.

**Figure 2.** The magnitude-redshift distribution of the new data (total galaxy sample) compared to earlier work. The three solid lines show the contours below which lie 5%, 50% and 95% of the galaxies in the no-evolution model described in Section 4.3.1.

**Figure 3.** Optical colours as a function of redshift for the $B < B_{70\%}$ galaxy sample, together with those of stars and unidentified objects to the left and right. Arrows represent $3\sigma$ upper limits when $R > 24$. Loci of some K-corrected non-evolving spectral types are shown.

**Figure 4.** The [OII] 3727 Å emission line equivalent width versus optical colour. Points with $W_\lambda = 0$ are where the object has a redshift but no [OII] — in this case the error bar represents the upper limit.

**Figure 5.** The redshift distribution of the new survey compared with the various galaxy evolution models described in the text: (a) The $B < B_{70\%}$ sample of 73 galaxies. (b) The 89% complete sub-sample of 30 galaxies.

*A faint galaxy redshift survey to B = 24*     9TABLE 1
PHOTOMETRIC OBSERVATIONS

| Field | R.A. (1950) Dec. | | Area† | Time | Seeing | $B_{lim}$ | N* |
|---|---|---|---|---|---|---|---|
| 03Z3 | 03 39 35 | −00 08 00 | 66.9 | 5000s | 1.6 | 24.1 | 504 |
| 10Z2 | 10 44 01 | +00 05 26 | 45.1 | 5000s | 2.1 | 24.5 | 110 |
| 13Z2 | 13 41 43 | +00 06 55 | 51.4 | 4200s | 1.6 | 24.5 | 170 |
| 22Z3 | 22 02 26 | −18 49 50 | 46.4 | 5000s | 1.5 | 24.1 | 218 |

* for $22.5 < B < 24$
† in arcmin$^2$

TABLE 2
SPECTROSCOPIC OBSERVATIONS

| Mask | R.A. (1950) Dec. | | Slits | Time | Seeing | IDs | $B_{70\%}$ | Slits‡ | IDs‡ | $A_{eff}$¶ |
|---|---|---|---|---|---|---|---|---|---|---|
| 03Z3_A | 03 39 36.0 | −00 09 05 | 30 | 14400s | 1.1″ | 21 | 24.0 | 30 | 21 | 5.68 |
| 10Z2_A | 10 43 58.0 | +00 05 40 | 20 | 13500s | 1.1″ | 15 | 24.0 | 20 | 15 | 3.90 |
| (10Z2_B)* | 10 43 58.0 | +00 05 40 | 20 | 15000s | 3.0″ | 10 | 23.5 | 11 | 8 | 4.06 |
| 10Z2_B | 10 43 58.0 | +00 05 40 | 20 | 10250s | 1.2″ | 16 | 24.0 | 20 | 16 | 3.40 |
| 10HST1 | 10 43 58.0 | +00 05 40 | 10† | 18000s | 1.5–2.0″ | 7 | 24.0 | 10 | 7 | 1.79 |
| 13Z2_A | 13 41 42.1 | +00 07 11 | 24 | 9000s | 1.2″ | 12 | 23.3 | 13 | 9 | 7.69 |
| 13HST1 | 13 41 42.1 | +00 07 11 | 19† | 19800s | 1.3–2.5″ | 10 | 23.5 | 12 | 9 | 4.50 |
| 22Z3_A | 22 02 27.5 | −18 49 50 | 34 | 19800s | 1.3″ | 13 | 23.0 | 6 | 4 | 6.69 |

* Earlier observation
† + extra objects with $B < 22.5$ — see Table 4 for details
‡ $B < B_{70\%}$ sample
¶ Effective area in arcmin$^2$

TABLE 4A
STATISTICS OF $n(z)$ MODELS FOR $B < B_{70\%}$ SAMPLE.

| | $P_{KS}$ | $z_{MED}$ | $P_{MED}$ | $f_{0.7}$* | $P_{0.7}$* |
|---|---|---|---|---|---|
| Data | — | 0.46 | — | 0.12 | — |
| No Evolution | $2 \times 10^{-3}$ | 0.53 | 0.88 | 0.26 | 1.00 |
| $b = 2, Q = 0$ | $3 \times 10^{-19}$ | 0.83 | $< 10^{-6}$ | 0.61 | $3 \times 10^{-6}$ |
| $b = 4, Q = 0$ | $9 \times 10^{-37}$ | 1.39 | $< 10^{-6}$ | 0.84 | $< 10^{-6}$ |
| $b = 3, Q = 4$ | $2 \times 10^{-2}$ | 0.51 | 0.95 | 0.21 | 0.99 |
| Koo *et al.* 1993 | $1 \times 10^{-5}$ | 0.56 | 0.50 | 0.36 | 0.73 |

* See text for definitions

TABLE 4B
STATISTICS OF $n(z)$ MODELS FOR 89% COMPLETE SUB-SAMPLE.

| | $P_{KS}$ | $z_{MED}$ | $P_{MED}$ | $f_{0.7}$* | $P_{0.7}$* |
|---|---|---|---|---|---|
| Data | — | 0.46 | — | 0.13 | — |
| No Evolution | 0.45 | 0.49 | 0.17 | 0.19 | 0.62 |
| $b = 2, Q = 0$ | $7 \times 10^{-7}$ | 0.76 | $8 \times 10^{-5}$ | 0.55 | $2 \times 10^{-4}$ |
| $b = 4, Q = 0$ | $1 \times 10^{-15}$ | 1.31 | $< 10^{-6}$ | 0.82 | $< 10^{-6}$ |
| $b = 3, Q = 4$ | 0.35 | 0.47 | 0.22 | 0.14 | 0.46 |
| Koo *et al.* 1993 | $1 \times 10^{-2}$ | 0.56 | $5 \times 10^{-2}$ | 0.36 | 0.10 |

* See text for definitions

TABLE 3
The redshift catalogue

| Mask | Slit | ID | R.A. (1950) | Dec. | $B$ | $R$ | $z$ | Ty | $W_\lambda$(Å) | Q | Comments |
|---|---|---|---|---|---|---|---|---|---|---|---|
| 03z3_A | 1 | 03.505 | 03 39 39.34 | −00 13 43.31 | 23.26 ± 0.07 | 21.38 ± 0.02 | 0.617 | EA | 17 ± 2 | 1 | 1 OII,H+K |
| 03z3_A | 2 | 03.524 | 03 39 37.88 | −00 13 21.73 | 23.34 ± 0.08 | 22.23 ± 0.04 | 0.616 | EA | 33 ± 4 | 1 | 1 OII,H+K,Balmer,G,H$\beta$+? |
| 03z3_A | 3 | 03.944 | 03 39 34.05 | −00 13 07.06 | 22.86 ± 0.06 | No Data | 0.303 | A | ≤ 4 | 1 | 1 H+K,Mgb |
| 03z3_A | 4 | 03.572 | 03 39 38.46 | −00 12 39.43 | 23.81 ± 0.13 | 23.30 ± 0.12 | UnID | — | — | 4 | 4 Missing? |
| 03z3_A | 5 | 03.595 | 03 39 40.90 | −00 12 19.18 | 22.68 ± 0.04 | 21.15 ± 0.02 | 0.594 | EA | 10 ± 2 | 1 | 1 OII,H+K,H$\delta$-,MgII- |
| 03z3_A | 6 | 03.608 | 03 39 41.49 | −00 12 04.53 | 23.87 ± 0.12 | 23.40 ± 0.12 | 0.893 | E | 63 ± 12 | 2 | 2 ?OII |
| 03z3_A | 7 | 03.619 | 03 39 36.03 | −00 11 50.49 | 23.18 ± 0.07 | 22.38 ± 0.05 | 0.750 | E | 41 ± 3 | 2 | 2 ?OII |
| 03z3_A | 8 | 03.644 | 03 39 38.76 | −00 11 27.05 | 23.94 ± 0.13 | 21.96 ± 0.03 | 0.000 | S | — | 1 | 1 M star |
| 03z3_A | 9 | 03.668 | 03 39 40.47 | −00 11 09.96 | 23.19 ± 0.07 | 22.27 ± 0.04 | 0.432 | E | 84 ± 3 | 1 | 1 OII,?H$\gamma$+,H$\beta$+,OIII,OIII |
| 03z3_A | 10 | 03.687 | 03 39 39.49 | −00 10 50.83 | 22.83 ± 0.05 | 21.38 ± 0.02 | 0.302 | EA | 18 ± 2 | 1 | 1 OII,H+K,H$\delta$-,?G,?H$\beta$+,Mgb |
| 03z3_A | 11 | 03.723 | 03 39 35.93 | −00 10 21.58 | 22.51 ± 0.04 | 21.58 ± 0.02 | 0.315 | EA | 54 ± 5 | 1 | 1 OII,H+K,H$\beta$+,OIII,OIII |
| 03z3_A | 12 | 03.740 | 03 39 38.47 | −00 10 06.59 | 23.67 ± 0.04 | 22.75 ± 0.07 | 0.432 | E | 54 ± 7 | 1 | 1 OII,OIII,OIII |
| 03z3_A | 13 | 03.763 | 03 39 42.52 | −00 09 41.84 | 23.90 ± 0.12 | 22.49 ± 0.05 | UnID | — | — | 4 | 4 Missing? |
| 03z3_A | 14 | 03.773 | 03 39 43.78 | −00 09 27.67 | 22.87 ± 0.05 | 21.96 ± 0.03 | 0.375 | EA | 62 ± 6 | 1 | 1 OII,?H$\beta$+,?OIII,OIII |
| 03z3_A | 15 | 03.788 | 03 39 42.75 | −00 09 11.49 | 23.91 ± 0.13 | 22.93 ± 0.08 | 0.540 | E | 34 ± 5 | 2 | 2 OII,OIII, break |
| 03z3_A | 16 | 03.803 | 03 39 41.62 | −00 08 58.35 | 23.77 ± 0.12 | 23.63 ± 0.16 | UnID | — | — | 3 | 3 Weak |
| 03z3_A | 17 | 03.820 | 03 39 43.79 | −00 08 39.03 | 23.86 ± 0.13 | 23.07 ± 0.10 | 0.487 | E | 31 ± 4 | 2 | 2 ?OII |
| 03z3_A | 18 | 03.842 | 03 39 39.25 | −00 08 14.04 | 23.60 ± 0.11 | Too faint | 0.000 | S | — | 1 | 1 Star |
| 03z3_A | 19 | 03.023 | 03 39 32.71 | −00 07 46.31 | 23.97 ± 0.13 | 23.03 ± 0.10 | UnID | — | — | 3 | 3 Weak |
| 03z3_A | 20 | 03.033 | 03 39 26.51 | −00 07 32.81 | 23.99 ± 0.12 | 22.65 ± 0.07 | 0.646 | E | 69 ± 5 | 2 | 2 ?OII |
| 03z3_A | 21 | 03.059 | 03 39 30.40 | −00 07 19.60 | 23.22 ± 0.06 | 22.15 ± 0.05 | 0.596 | E | 60 ± 2 | 2 | 2 ?OII,?H$\delta$+- |
| 03z3_A | 22 | 03.096 | 03 39 26.01 | −00 06 42.00 | 23.85 ± 0.10 | 23.11 ± 0.10 | 0.581 | E | 85 ± 3 | 2 | 2 ?OII |
| 03z3_A | 23 | 03.114 | 03 39 28.94 | −00 06 26.18 | 23.32 ± 0.06 | 22.89 ± 0.08 | UnID | — | — | 3 | 3 Featureless |
| 03z3_A | 24 | 03.132 | 03 39 29.58 | −00 06 07.44 | 23.91 ± 0.11 | 22.75 ± 0.07 | 0.288 | EA | 37 ± 4 | 1 | 1 OII,OIII |
| 03z3_A | 25 | 03.155 | 03 39 31.40 | −00 05 47.00 | 23.60 ± 0.08 | No Data | UnID | — | — | 4 | 4 Missing? |
| 03z3_A | 26 | 03.202 | 03 39 29.23 | −00 05 19.89 | 23.51 ± 0.07 | 21.87 ± 0.03 | 0.746 | E | 12 ± 2 | 2 | 2 ?OII,??H+K (?OII=MgII at z=1.324?) |
| 03z3_A | 27 | 03.219 | 03 39 30.42 | −00 05 05.51 | 23.59 ± 0.08 | 22.48 ± 0.06 | 0.601 | E | 37 ± 4 | 2 | 2 ?OII |
| 03z3_A | 28 | 03.241 | 03 39 31.36 | −00 04 48.28 | 23.68 ± 0.08 | 23.03 ± 0.09 | UnID | — | — | 3 | 3 Weak |
| 03z3_A | 29 | 03.279 | 03 39 31.46 | −00 04 20.76 | 23.97 ± 0.12 | 23.79 ± 0.18 | UnID | — | — | 4 | 4 Missing? |
| 03z3_A | 30 | 03.307 | 03 39 31.40 | −00 04 05.83 | 23.28 ± 0.07 | 22.29 ± 0.05 | UnID | — | — | 3 | 3 Featureless |
| 10z2_A | 1 | 10.205 | 10 43 58.32 | 00 01 47.32 | 23.52 ± 0.09 | 23.50 ± 0.21 | UnID | — | — | 3 | 3 Featureless |
| 10z2_A | 2 | 10.227 | 10 43 53.70 | 00 02 27.39 | 23.98 ± 0.11 | 23.55 ± 0.19 | UnID | — | — | 3 | 3 Weak |
| 10z2_A | 3 | 10.233 | 10 43 51.83 | 00 02 40.73 | 22.72 ± 0.04 | 20.99 ± 0.02 | 0.307 | EA | 20 ± 2 | 1 | 1 OII,H+K,H$\delta$-,H$\gamma$-,?OIII |
| 10z2_A | 4 | 10.262 | 10 43 52.62 | 00 03 25.60 | 23.77 ± 0.10 | 21.70 ± 0.03 | 0.294 | EA | 12 ± 4 | 1 | 1 OII,FeI,H+K,?OIII |
| 10z2_A | 5 | 10.250 | 10 44 06.38 | 00 03 04.36 | 23.56 ± 0.08 | 22.99 ± 0.12 | UnID | — | — | 3 | 3 Featureless |
| 10z2_A | 6 | 10.279 | 10 44 00.14 | 00 03 47.02 | 22.54 ± 0.03 | 22.27 ± 0.06 | 0.634 | E | 55 ± 1 | 1 | 1 OII,H$\beta$+,OIII,OIII |
| 10z2_A | 7 | 10.288 | 10 43 59.13 | 00 04 07.80 | 23.37 ± 0.07 | 23.08 ± 0.12 | 1.108 | EA | 65 ± 10 | 1 | 1 OII,MgI-,MgII-,?FeII- |
| 10z2_A | 8 | 10.313 | 10 44 06.18 | 00 04 55.71 | 23.88 ± 0.11 | 21.20 ± 0.02 | 0.000 | S | — | 1 | 1 M star |
| 10z2_A | 9 | 10.328 | 10 44 05.38 | 00 05 29.63 | 23.16 ± 0.06 | 21.25 ± 0.02 | 0.207 | A | 19 ± 6 | 1 | 1 H+K,G,Mgb,NaD |
| 10z2_A | 10 | 10.022 | 10 44 01.49 | 00 05 48.04 | 23.96 ± 0.15 | Too faint | 0.924 | E | 17 ± 8 | 2 | 2 ?OII,?MgII- |
| 10z2_A | 11 | 10.031 | 10 43 52.15 | 00 06 04.33 | 23.52 ± 0.10 | 22.45 ± 0.07 | 0.278 | EA | 41 ± 4 | 1 | 1 OII,H$\beta$+,OIII,OIII |
| 10z2_A | 12 | 10.048 | 10 44 10.82 | 00 06 34.51 | 22.93 ± 0.06 | 20.78 ± 0.04 | 0.276 | EA | 27 ± 2 | 1 | 1 OII,H+K,H$\beta$+,OIII,OIII |
| 10z2_A | 13 | 10.066 | 10 44 04.19 | 00 07 00.49 | 23.30 ± 0.08 | 21.80 ± 0.05 | 0.368 | EA | 15 ± 2 | 1 | 1 OII,H+K,H$\beta$+,OIII,Mgb |
| 10z2_A | 14 | 10.080 | 10 44 03.35 | 00 07 38.20 | 22.78 ± 0.05 | 21.93 ± 0.05 | 0.621 | E | 37 ± 1 | 1 | 1 OII,?H$\delta$-,?H$\beta$+,OIII |
| 10z2_A | 15 | 10.088 | 10 43 56.16 | 00 07 56.80 | 22.70 ± 0.04 | 21.60 ± 0.04 | 0.177 | EA | ≤ 8 | 1 | 1 H+K,G,H$\alpha$+ |
| 10z2_A | 16 | 10.073 | 10 43 51.93 | 00 07 16.95 | 23.83 ± 0.12 | Too faint | UnID | — | — | 4 | 4 Missing? |
| 10z2_A | 17 | 10.109 | 10 43 59.45 | 00 08 39.33 | 23.66 ± 0.11 | 20.74 ± 0.02 | 0.492 | A | — | 1 | 1 ?OII,H+K,H$\delta$-,G,H$\gamma$- |
| 10z2_A | 18 | 10.120 | 10 43 55.03 | 00 08 57.26 | 23.51 ± 0.09 | 21.11 ± 0.02 | 0.579 | EA | 6 ± 1 | 1 | 1 ?OII,H+K,FeI,G (OII on NS) |
| 10z2_A | 19 | 10.131 | 10 44 11.93 | 00 09 20.17 | 23.37 ± 0.12 | 22.75 ± 0.09 | UnID | — | — | 4 | 4 Missing? |
| 10z2_A | 20 | 10.301 | 10 44 02.95 | 00 04 36.28 | 23.51 ± 0.08 | 21.97 ± 0.04 | 0.323 | EA | 25 ± 5 | 1 | 1 OII,OIII,OIII |
| 10z2_B | 1 | 10.206 | 10 44 02.71 | 00 01 48.30 | 23.48 ± 0.08 | 22.69 ± 0.09 | 1.599 | Q | — | 1 | 1 QSO? - MgII,CIII],CIV |
| 10z2_B | 2 | 10.223 | 10 43 55.52 | 00 02 23.66 | 23.99 ± 0.12 | 23.61 ± 0.19 | 0.296 | E | 58 ± 8 | 1 | 1 OII,H$\beta$,OII, ?H+K |
| 10z2_B | 3 | 10.236 | 10 44 05.91 | 00 02 43.60 | 23.81 ± 0.10 | Too faint | UnID | — | — | 4 | 4 Missing? |
| 10z2_B | 4 | 10.248 | 10 44 00.02 | 00 03 00.54 | 23.23 ± 0.06 | 21.07 ± 0.02 | 0.314 | A | ≤ 3 | 1 | 1 H+K |
| 10z2_B | 5 | 10.260 | 10 43 56.37 | 00 03 20.87 | 23.49 ± 0.08 | 21.95 ± 0.05 | 0.563 | EA | 17 ± 2 | 1 | 1 OII,H+K,Balmer |
| 10z2_B | 6 | 10.277 | 10 44 11.03 | 00 03 44.77 | 23.10 ± 0.05 | 22.34 ± 0.06 | UnID | — | — | 3 | 3 Incorrect mag? |
| 10z2_B | 7 | 10.286 | 10 44 02.43 | 00 04 03.01 | 22.65 ± 0.04 | 21.43 ± 0.03 | 0.559 | EA | 33 ± 2 | 1 | 1 OII,H+K,Balmer |
| 10z2_B | 8 | 10.300 | 10 44 00.65 | 00 04 36.37 | 23.96 ± 0.12 | 21.21 ± 0.02 | 0.000 | S | — | 1 | 1 M star (flux too low for mag) |
| 10z2_B | 9 | 10.315 | 10 44 02.62 | 00 05 00.55 | 23.98 ± 0.12 | Too faint | UnID | — | — | 3 | 3 Weak |
| 10z2_B | 10 | 10.330 | 10 44 03.23 | 00 05 31.41 | 23.70 ± 0.09 | 22.26 ± 0.06 | 0.324 | E | 50 ± 3 | 1 | 1 OII,H$\beta$,OIII,?H+K |
| 10z2_B | 11 | 10.025 | 10 43 59.13 | 00 05 55.72 | 22.68 ± 0.04 | 23.62 ± 0.24 | 2.749 | Q | — | 1 | 1 Ly$\alpha$,CIV,?CIII] |
| 10z2_B | 12 | 10.036 | 10 43 59.43 | 00 06 16.07 | 23.02 ± 0.06 | 21.51 ± 0.03 | 0.478 | EA | 30 ± 4 | 1 | 1 OII,OIII,H+K,Balmer |
| 10z2_B | 13 | 10.061 | 10 44 01.02 | 00 06 52.30 | 23.14 ± 0.07 | 20.92 ± 0.03 | 0.384 | A | ≤ 3 | 1 | 1 H+K,G,H$\beta$,Mgb |
| 10z2_B | 14 | 10.047 | 10 43 53.94 | 00 06 34.04 | 23.87 ± 0.13 | 23.98 ± 0.38 | 0.199 | E | 26 ± 4 | 1 | 1 OII,H$\beta$,OIII,?H+K,?Balmer |
| 10z2_B | 15 | 10.071 | 10 43 56.75 | 00 07 10.47 | 22.84 ± 0.05 | 20.50 ± 0.01 | 0.476 | EA | 10 ± 2 | 1 | 1 OII,H+K,G,Balmer |
| 10z2_B | 16 | 10.077 | 10 43 55.39 | 00 07 33.06 | 23.70 ± 0.11 | 21.48 ± 0.03 | 0.436 | A | ≤ 2 | 1 | 1 H+K,G,Balmer |
| 10z2_B | 17 | 10.086 | 10 43 57.58 | 00 07 50.52 | 23.73 ± 0.12 | 23.93 ± 0.31 | UnID | — | — | 3 | 3 Weak |
| 10z2_B | 18 | 10.107 | 10 44 07.20 | 00 08 38.23 | 23.87 ± 0.14 | 22.10 ± 0.06 | 0.448 | E | 55 ± 5 | 1 | 1 OII,H$\beta$,OIII |
| 10z2_B | 19 | 10.122 | 10 44 02.20 | 00 09 00.41 | 23.11 ± 0.06 | 22.03 ± 0.06 | 0.724 | EA | 30 ± 2 | 2 | 2 OII,Hz,K,(H in sky abs?) |
| 10z2_B | 20 | 10.130 | 10 44 08.65 | 00 09 19.88 | 23.19 ± 0.07 | 20.87 ± 0.02 | 0.456 | EA | 8 ± 2 | 1 | 1 OII,H+K,Balmer,?OIII |

| Mask | Slit | ID | R.A. (1950) | Dec. | $B$ | $R$ | $z$ | Ty | $W_\lambda$(Å) | Q | Comments |
|---|---|---|---|---|---|---|---|---|---|---|---|
| 10HST1 | 1 | 10.204 | 10 43 57.17 | 00 01 45.04 | 23.10 ± 0.06 | 22.61 ± 0.09 | 0.742 | EA | 42 ± 3 | 1 | 1 O II, Fe II, break? |
| 10HST1 | 2 | 10.218 | 10 43 57.36 | 00 01 57.47 | 20.72 ± 0.00 | 18.93 ± 0.00 | 0.097 | A | ≤ 5 | 1 | 1 H+K, G, Hβ-, NaD |
| 10HST1 | 3 | 10.222 | 10 43 57.69 | 00 02 15.95 | 22.39 ± 0.03 | 22.75 ± 0.10 | 1.999 | Q | — | 1 | 1 QSO  C IV, C III, no Lα? |
| 10HST1 | 4 | 10.227 | 10 43 53.70 | 00 02 27.39 | 23.98 ± 0.11 | 23.55 ± 0.19 | UnID | — | — | 3 | 3 Weak |
| 10HST1 | 5 | 10.235 | 10 43 55.01 | 00 02 42.11 | 21.95 ± 0.02 | 19.65 ± 0.00 | 0.000 | S | — | 1 | 1 M star |
| 10HST1 | 6 | 10.249 | 10 43 51.44 | 00 03 02.44 | 23.50 ± 0.08 | 22.35 ± 0.06 | 0.466 | EA | 8 ± 4 | 1 | 1 O II, O III, H+K |
| 10HST1 | 7 | 10.255 | 10 43 55.59 | 00 03 15.08 | 22.52 ± 0.04 | 21.44 ± 0.03 | 0.149 | EA | 42 ± 5 | 1 | 1 O II, O III, Hα, H+K |
| 10HST1 | 8 | 10.273 | 10 43 56.99 | 00 03 35.55 | 22.47 ± 0.03 | 19.47 ± 0.00 | 0.435 | EA | 3 ± 1 | 1 | 1 O II, H+K, G, |
| 10HST1 | 9 | 10.315 | 10 44 02.62 | 00 05 00.55 | 23.98 ± 0.12 | Too faint | UnID | — | — | 4 | 4 Missing |
| 10HST1 | 10 | 10.332 | 10 44 01.27 | 00 05 38.72 | 22.78 ± 0.04 | 20.60 ± 0.01 | 0.000 | S | — | 1 | 1 late-type star |
| 10HST1 | 11 | 10.028 | 10 44 03.88 | 00 05 59.00 | 21.89 ± 0.02 | 20.64 ± 0.01 | 0.582 | EA | 8 ± 1 | 1 | 1 O II, O III, Balmer, |
| 10HST1 | 12 | 10.032 | 10 44 01.53 | 00 06 09.78 | 23.94 ± 0.14 | Too faint | UnID | — | — | 4 | 4 Missing |
| 10HST1 | 13 | 10.040 | 10 43 57.82 | 00 06 21.31 | 23.73 ± 0.12 | 20.96 ± 0.02 | 0.476 | A | ≤ 7 | 1 | 1 H+K, G, Hβ-, Mgb |
| 10HST1 | 14 | 10.051 | 10 44 01.99 | 00 06 35.95 | 22.84 ± 0.05 | 22.65 ± 0.10 | 0.081 | E | 20 ± 9 | 1 | 1 Hα, O III |
| 10HST1 | 15 | 10.064 | 10 44 04.88 | 00 06 54.81 | 22.11 ± 0.03 | 19.80 ± 0.00 | 0.000 | S | — | 1 | 1 M star |
| 10HST1 | 16 | 10.086 | 10 43 57.58 | 00 07 50.52 | 23.73 ± 0.12 | 23.93 ± 0.31 | 0.758 | E | 24 ± 2 | 2 | 2 O II only |
| 10HST1 | 17 | 10.092 | 10 44 01.26 | 00 08 09.05 | 22.16 ± 0.03 | 21.18 ± 0.03 | UnID | — | — | 3 | 3 Featureless |
| 10HST1 | 18 | 10.105 | 10 43 57.66 | 00 08 25.41 | 20.02 ± 0.00 | 17.83 ± 0.00 | 0.000 | S | — | 1 | 1 M star |
| 10HST1 | 19 | 10.116 | 10 43 55.84 | 00 08 50.06 | 21.73 ± 0.02 | 21.51 ± 0.04 | 1.256 | Q | — | 1 | 1 QSO |
| 10HST1 | 20 | 10.126 | 10 43 58.72 | 00 09 05.44 | 22.30 ± 0.03 | 21.69 ± 0.05 | 0.000 | S | — | 1 | 1 early-type star |
| 13z2_A | 1 | 13.304 | 13 41 42.70 | 00 02 49.73 | 23.66 ± 0.09 | 23.66 ± 0.09 | UnID | — | — | 4 | 4 Missing? |
| 13z2_A | 2 | 13.311 | 13 41 41.44 | 00 03 05.87 | 23.38 ± 0.06 | 22.04 ± 0.02 | UnID | — | — | 3 | 3 Weak |
| 13z2_A | 3 | 13.323 | 13 41 38.73 | 00 03 24.07 | 23.16 ± 0.05 | 21.29 ± 0.03 | 0.385 | EA | 19 ± 2 | 1 | 1 O II, H+K, Fe I, Hδ- |
| 13z2_A | 4 | 13.347 | 13 41 48.56 | 00 03 45.36 | 23.09 ± 0.05 | 22.32 ± 0.03 | UnID | — | — | 3 | 3 Weak |
| 13z2_A | 5 | 13.370 | 13 41 35.39 | 00 04 11.08 | 23.78 ± 0.09 | 22.49 ± 0.03 | UnID | — | — | 3 | 3 Weak (?em@4775,6745?) |
| 13z2_A | 6 | 13.384 | 13 41 39.08 | 00 04 29.73 | 23.65 ± 0.09 | 22.58 ± 0.03 | UnID | — | — | 3 | 3 Weak |
| 13z2_A | 7 | 13.402 | 13 41 45.60 | 00 04 49.09 | 23.24 ± 0.06 | 22.11 ± 0.02 | 0.830 | E | 44 ± 4 | 2 | 2 O II |
| 13z2_A | 8 | 13.420 | 13 41 34.97 | 00 05 14.33 | 23.85 ± 0.10 | 22.07 ± 0.03 | UnID | — | — | 3 | 3 Weak |
| 13z2_A | 9 | 13.444 | 13 41 39.28 | 00 05 38.11 | 23.21 ± 0.06 | 22.60 ± 0.04 | UnID | — | — | 3 | 3 Weak |
| 13z2_A | 10 | 13.465 | 13 41 40.75 | 00 05 59.44 | 23.23 ± 0.06 | 21.90 ± 0.02 | 0.556 | EA | 14 ± 3 | 1 | 1 O II, H+K, Hδ-, Hγ- |
| 13z2_A | 11 | 13.480 | 13 41 43.00 | 00 06 14.39 | 22.83 ± 0.04 | 21.49 ± 0.01 | 0.556 | EA | 34 ± 5 | 2 | 2 ?O II |
| 13z2_A | 12 | 13.504 | 13 41 48.79 | 00 06 48.80 | 23.40 ± 0.06 | 22.05 ± 0.02 | UnID | — | — | 3 | 3 Weak |
| 13z2_A | 13 | 13.517 | 13 41 40.95 | 00 07 06.57 | 23.94 ± 0.10 | 22.51 ± 0.03 | 0.462 | E | 36 ± 7 | 2 | 2 ?O II |
| 13z2_A | 14 | 13.016 | 13 41 45.13 | 00 07 28.62 | 22.82 ± 0.03 | 21.89 ± 0.01 | UnID | — | — | 3 | 3 Featureless |
| 13z2_A | 15 | 13.027 | 13 41 42.29 | 00 07 50.95 | 22.86 ± 0.03 | 22.48 ± 0.02 | 0.089 | E | 41 ± 7 | 1 | 1 Hβ+, O III, O III, Hα (?O II?, ?EW?) |
| 13z2_A | 16 | 13.038 | 13 41 44.33 | 00 08 10.76 | 22.88 ± 0.04 | 21.40 ± 0.00 | 0.424 | EA | 12 ± 2 | 1 | 1 O II, H+K, G |
| 13z2_A | 17 | 13.056 | 13 41 38.74 | 00 08 38.20 | 22.70 ± 0.03 | 21.33 ± 0.00 | 0.556 | EA | 32 ± 2 | 2 | 2 ?O II |
| 13z2_A | 18 | 13.078 | 13 41 42.95 | 00 08 57.78 | 23.33 ± 0.05 | 21.85 ± 0.01 | UnID | — | — | 3 | 3 Weak |
| 13z2_A | 19 | 13.087 | 13 41 35.40 | 00 09 16.17 | 23.17 ± 0.04 | 20.71 ± 0.00 | 0.359 | A | ≤ 2 | 1 | 1 H+K, G |
| 13z2_A | 20 | 13.106 | 13 41 44.15 | 00 09 36.95 | 22.92 ± 0.04 | 21.66 ± 0.01 | 0.187 | E | 59 ± 14 | 1 | 1 O II, Hβ+, O III, Hα |
| 13z2_A | 21 | 13.123 | 13 41 42.76 | 00 09 57.38 | 23.45 ± 0.05 | 22.11 ± 0.02 | 0.536 | E | 32 ± 3 | 2 | 2 ?O II |
| 13z2_A | 22 | 13.160 | 13 41 47.41 | 00 10 43.91 | 23.53 ± 0.06 | 21.96 ± 0.02 | 0.335 | E | 35 ± 4 | 1 | 1 O II, O III |
| 13z2_A | 23 | 13.177 | 13 41 52.15 | 00 11 05.46 | 23.02 ± 0.04 | 22.15 ± 0.02 | UnID | — | — | 3 | 3 Weak |
| 13z2_A | 24 | 13.190 | 13 41 44.30 | 00 11 27.96 | 23.76 ± 0.07 | 21.40 ± 0.00 | UnID | — | — | 3 | 3 Weak |
| 13HST1 | 1 | 13.311 | 13 41 41.44 | 00 03 05.87 | 23.38 ± 0.06 | 22.04 ± 0.02 | 0.458 | EA | 21 ± 5 | 2 | 2 O II, H+K |
| 13HST1 | 2 | 13.325 | 13 41 40.71 | 00 03 25.88 | 23.40 ± 0.07 | 20.70 ± 0.00 | 0.000 | S | — | 1 | 1 M star |
| 13HST1 | 3 | 13.344 | 13 41 38.97 | 00 03 41.64 | 23.24 ± 0.06 | 22.59 ± 0.04 | 0.452 | A | ≤ 117 | 2 | 2 H+K, |
| 13HST1 | 4 | 13.358 | 13 41 48.83 | 00 03 57.17 | 21.99 ± 0.02 | 21.02 ± 0.00 | UnID | — | — | 3 | 3 Weak |
| 13HST1 | 5 | 13.367 | 13 41 39.47 | 00 04 07.10 | 23.75 ± 0.10 | 22.60 ± 0.04 | UnID | — | — | 4 | 4 Missing |
| 13HST1 | 6 | 13.378 | 13 41 47.15 | 00 04 21.71 | 22.67 ± 0.03 | 21.73 ± 0.02 | UnID | — | — | 3 | 3 Weak |
| 13HST1 | 7 | 13.388 | 13 41 47.53 | 00 04 36.43 | 22.75 ± 0.04 | 20.55 ± 0.00 | 0.443 | A | ≤ 3 | 1 | 1 H+K, G |
| 13HST1 | 8 | 13.400 | 13 41 45.16 | 00 04 49.14 | 21.87 ± 0.02 | 20.57 ± 0.00 | 0.830 | E | 19 ± 1 | 1 | 1 O II |
| 13HST1 | 9 | 13.417 | 13 41 40.97 | 00 05 08.24 | 22.74 ± 0.04 | 20.43 ± 0.00 | 0.283 | A | ≤ 6 | 1 | 1 H+K, G, Hβ-, Mgb |
| 13HST1 | 10 | 13.452 | 13 41 43.26 | 00 05 40.86 | 22.00 ± 0.02 | 20.72 ± 0.00 | 0.451 | A | ≤ 4 | 2 | 2 H+K, G |
| 13HST1 | 11 | 13.460 | 13 41 44.21 | 00 05 54.13 | 22.61 ± 0.03 | 21.86 ± 0.02 | UnID | — | — | 3 | 3 Weak |
| 13HST1 | 12 | 13.469 | 13 41 42.38 | 00 06 03.46 | 23.07 ± 0.05 | 21.25 ± 0.01 | 0.493 | EA | 13 ± 1 | 1 | 1 O II, H+K, O III |
| 13HST1 | 13 | 13.484 | 13 41 45.37 | 00 06 24.56 | 20.58 ± 0.00 | 18.13 ± 0.00 | 0.000 | S | — | 1 | 1 M star |
| 13HST1 | 14 | 13.492 | 13 41 40.63 | 00 06 37.63 | 23.62 ± 0.08 | 22.29 ± 0.03 | UnID | — | — | 4 | 4 Missing |
| 13HST1 | 15 | 13.510 | 13 41 43.50 | 00 06 55.26 | 23.01 ± 0.05 | 20.89 ± 0.00 | 0.566 | EA | 7 ± 1 | 1 | 1 O II, H+K, Hδ-, Hγ-, Hβ- |
| 13HST1 | 16 | 13.519 | 13 41 41.84 | 00 07 06.81 | 22.39 ± 0.03 | 20.75 ± 0.00 | 0.278 | EA | 8 ± 3 | 1 | 1 O II?, H+K, Hβ-, Mgb, NaD |
| 13HST1 | 17 | 13.012 | 13 41 42.71 | 00 07 22.29 | 22.09 ± 0.01 | 21.05 ± 0.00 | UnID | — | — | 3 | 3 Featureless |
| 13HST1 | 18 | 13.028 | 13 41 44.80 | 00 07 52.37 | 23.41 ± 0.05 | 22.01 ± 0.01 | 0.426 | EA | 4 ± 21 | 2 | 2 O II, O III, Balmer |
| 13HST1 | 19 | 13.034 | 13 41 42.78 | 00 08 04.94 | 23.73 ± 0.07 | 22.69 ± 0.03 | UnID | — | — | 3 | 3 Featureless |
| 13HST1 | 20 | 13.079 | 13 41 42.11 | 00 08 59.55 | 21.78 ± 0.01 | 19.65 ± 0.00 | 0.279 | A | ≤ 2 | 1 | 1 H+K, G, Hβ-, NaD, Mgb |
| 13HST1 | 21 | 13.091 | 13 41 43.02 | 00 09 21.15 | 23.64 ± 0.06 | 22.46 ± 0.02 | UnID | — | — | 3 | 3 Weak |
| 13HST1 | 22 | 13.098 | 13 41 43.19 | 00 09 32.00 | 23.22 ± 0.04 | 22.56 ± 0.03 | UnID | — | — | 3 | 3 Featureless |
| 13HST1 | 23 | 13.116 | 13 41 46.87 | 00 09 50.09 | 22.79 ± 0.03 | 20.82 ± 0.00 | 0.363 | EA | 7 ± 2 | 1 | 1 O II, H+K, Hβ- |
| 13HST1 | 24 | 13.131 | 13 41 47.17 | 00 10 02.93 | 21.70 ± 0.01 | 19.91 ± 0.00 | 0.326 | EA | 4 ± 1 | 1 | 1 O II, Hδ-, G, Hβ- |
| 13HST1 | 25 | 13.139 | 13 41 41.42 | 00 10 13.89 | 23.73 ± 0.07 | 22.89 ± 0.04 | 0.146 | E | 29 ± 6 | 1 | 1 O II, O III, Hα |
| 13HST1 | 26 | 13.151 | 13 41 41.17 | 00 10 23.60 | 20.62 ± 0.00 | 19.86 ± 0.00 | 0.148 | A | ≤ 9 | 1 | 1 H+K, G, Hβ- |
| 13HST1 | 27 | 13.161 | 13 41 46.57 | 00 10 39.68 | 20.18 ± 0.00 | 18.85 ± 0.00 | 0.088 | A | ≤ 3 | 1 | 1 H+K, G, Hβ-, Mgb, NaD |
| 13HST1 | 28 | 13.164 | 13 41 51.34 | 00 10 52.48 | 23.86 ± 0.08 | 22.16 ± 0.02 | UnID | — | — | 3 | 3 Weak |
| 13HST1 | 29 | 13.180 | 13 41 46.61 | 00 11 07.51 | 22.16 ± 0.02 | 20.14 ± 0.00 | 0.000 | S | — | 1 | 1 M star |
| 13HST1 | 30 | 13.186 | 13 41 48.61 | 00 11 19.60 | 23.71 ± 0.07 | 22.50 ± 0.03 | UnID | — | — | 3 | 3 Weak |
| 13HST1 | 31 | 13.191 | 13 41 46.84 | 00 11 29.32 | 21.93 ± 0.01 | 21.16 ± 0.00 | 0.242 | EA | 65 ± 4 | 1 | 1 O II, H+K, O III, Hα |

| Mask | Slit | ID | R.A. | (1950) | Dec. | $B$ | $R$ | $z$ | Ty | $W_\lambda$(Å) | $Q$ | Comments |
|---|---|---|---|---|---|---|---|---|---|---|---|---|
| 22z3_A | 1 | 22.303 | 22 02 21.04 | −18 54 37.33 | | 23.92 ± 0.12 | 23.11 ± 0.12 | UnID | — | — | 4 | 4 Missing? |
| 22z3_A | 2 | 22.310 | 22 02 19.43 | −18 54 23.82 | | 23.47 ± 0.09 | 22.43 ± 0.06 | UnID | — | — | 3 | 3 Weak |
| 22z3_A | 3 | 22.322 | 22 02 22.94 | −18 54 09.69 | | 22.82 ± 0.05 | Too faint | 0.469 | EA | 8 ± 2 | 1 | 1 OII, H+K, Hδ- |
| 22z3_A | 4 | 22.332 | 22 02 25.25 | −18 53 51.99 | | 23.19 ± 0.07 | Too faint | 0.824 | E | 64 ± 3 | 2 | 2 ?OII |
| 22z3_A | 5 | 22.342 | 22 02 24.45 | −18 53 19.59 | | 22.86 ± 0.05 | 21.53 ± 0.03 | 0.263 | E | 65 ± 3 | 1 | 1 OII, Hβ+, OIII, OIII |
| 22z3_A | 6 | 22.346 | 22 02 29.08 | −18 53 03.66 | | 23.53 ± 0.10 | 22.10 ± 0.05 | 0.622 | E | 29 ± 3 | 2 | 2 ?OII |
| 22z3_A | 7 | 22.358 | 22 02 29.25 | −18 52 42.81 | | 23.09 ± 0.06 | 21.81 ± 0.04 | UnID | — | — | 3 | 3 Featureless |
| 22z3_A | 8 | 22.370 | 22 02 32.18 | −18 52 28.59 | | 23.33 ± 0.08 | 23.28 ± 0.14 | UnID | — | — | 3 | 3 Weak |
| 22z3_A | 9 | 22.378 | 22 02 32.17 | −18 52 13.42 | | 23.42 ± 0.09 | 22.60 ± 0.08 | UnID | — | — | 3 | 3 Weak |
| 22z3_A | 10 | 22.381 | 22 02 23.61 | −18 51 58.63 | | 23.47 ± 0.09 | 21.24 ± 0.02 | 0.000 | S | — | 1 | 1 M star |
| 22z3_A | 11 | 22.390 | 22 02 22.24 | −18 51 43.21 | | 23.85 ± 0.13 | 22.91 ± 0.10 | UnID | — | — | 3 | 3 Weak |
| 22z3_A | 12 | 22.406 | 22 02 29.87 | −18 51 27.55 | | 23.64 ± 0.10 | 23.95 ± 0.26 | UnID | — | — | 4 | 4 Missing? |
| 22z3_A | 13 | 22.412 | 22 02 27.16 | −18 51 12.44 | | 23.84 ± 0.13 | 22.41 ± 0.06 | UnID | — | — | 3 | 3 Weak |
| 22z3_A | 14 | 22.442 | 22 02 33.68 | −18 50 34.87 | | 22.62 ± 0.05 | No Data | 0.549 | EA | 18 ± 2 | 2 | 2 ?OII, ?H+K, ?FeI |
| 22z3_A | 15 | 22.434 | 22 02 27.91 | −18 50 17.22 | | 23.32 ± 0.08 | 22.02 ± 0.05 | 0.621 | E | 45 ± 3 | 2 | 2 ?OII |
| 22z3_A | 16 | 22.438 | 22 02 21.41 | −18 50 03.81 | | 23.41 ± 0.09 | 22.17 ± 0.05 | 0.603 | EA | 46 ± 4 | 2 | 2 ?OII, ?H+K, ?FeI |
| 22z3_A | 17 | 22.002 | 22 02 26.17 | −18 49 50.57 | | 22.56 ± 0.05 | 21.40 ± 0.02 | 0.399 | E | 31 ± 3 | 1 | 1 OII, Hβ+, OIII, OIII |
| 22z3_A | 18 | 22.014 | 22 02 19.11 | −18 49 23.00 | | 23.35 ± 0.09 | No Data | UnID | — | — | 3 | 3 Weak |
| 22z3_A | 19 | 22.027 | 22 02 27.82 | −18 49 03.57 | | 23.08 ± 0.07 | 22.26 ± 0.05 | UnID | — | — | 3 | 3 Weak |
| 22z3_A | 20 | 22.038 | 22 02 27.04 | −18 48 51.37 | | 23.45 ± 0.10 | 22.35 ± 0.05 | UnID | — | — | 3 | 3 Weak |
| 22z3_A | 21 | 22.052 | 22 02 26.40 | −18 48 33.72 | | 23.35 ± 0.09 | 21.30 ± 0.02 | 0.000 | S | — | 1 | 1 M star |
| 22z3_A | 22 | 22.065 | 22 02 24.25 | −18 48 18.11 | | 23.98 ± 0.16 | 23.75 ± 0.21 | UnID | — | — | 4 | 4 Missing |
| 22z3_A | 23 | 22.075 | 22 02 22.83 | −18 48 02.50 | | 23.48 ± 0.10 | Too faint | UnID | — | — | 3 | 3 Weak |
| 22z3_A | 24 | 22.090 | 22 02 24.90 | −18 47 48.50 | | 23.67 ± 0.12 | 22.66 ± 0.07 | UnID | — | — | 3 | 3 Featureless |
| 22z3_A | 25 | 22.270 | 22 02 24.14 | −18 47 32.86 | | 23.90 ± 0.15 | Too faint | 0.769 | E | 70 ± 7 | 2 | 2 ?OII |
| 22z3_A | 26 | 22.108 | 22 02 27.65 | −18 47 18.77 | | 23.76 ± 0.14 | Too faint | UnID | — | — | 4 | 4 Missing? |
| 22z3_A | 27 | 22.126 | 22 02 29.13 | −18 47 00.30 | | 22.73 ± 0.05 | 22.00 ± 0.04 | UnID | — | — | 3 | 3 Featureless |
| 22z3_A | 28 | 22.144 | 22 02 28.01 | −18 46 38.02 | | 23.96 ± 0.16 | 23.35 ± 0.14 | UnID | — | — | 4 | 4 Missing? |
| 22z3_A | 29 | 22.159 | 22 02 27.52 | −18 46 22.02 | | 23.56 ± 0.11 | 22.92 ± 0.09 | UnID | — | — | 3 | 3 Weak |
| 22z3_A | 30 | 22.172 | 22 02 26.53 | −18 46 03.62 | | 23.80 ± 0.14 | 23.20 ± 0.12 | 1.067 | E | 69 ± 15 | 2 | 2 OII, ?MgI, ?MgII |
| 22z3_A | 31 | 22.181 | 22 02 31.39 | −18 45 52.36 | | 23.96 ± 0.17 | 23.61 ± 0.19 | UnID | — | — | 4 | 4 Missing? |
| 22z3_A | 32 | 22.190 | 22 02 26.23 | −18 45 19.02 | | 23.87 ± 0.14 | 22.99 ± 0.09 | UnID | — | — | 3 | 3 Weak |
| 22z3_A | 33 | 22.205 | 22 02 25.81 | −18 44 58.86 | | 23.89 ± 0.14 | 22.34 ± 0.05 | 0.300 | E | 21 ± 6 | 1 | 1 OII, Hβ+, OIII, OIII |
| 22z3_A | 34 | 22.216 | 22 02 22.45 | −18 44 43.40 | | 22.86 ± 0.06 | 22.48 ± 0.06 | UnID | — | — | 3 | 3 Weak |

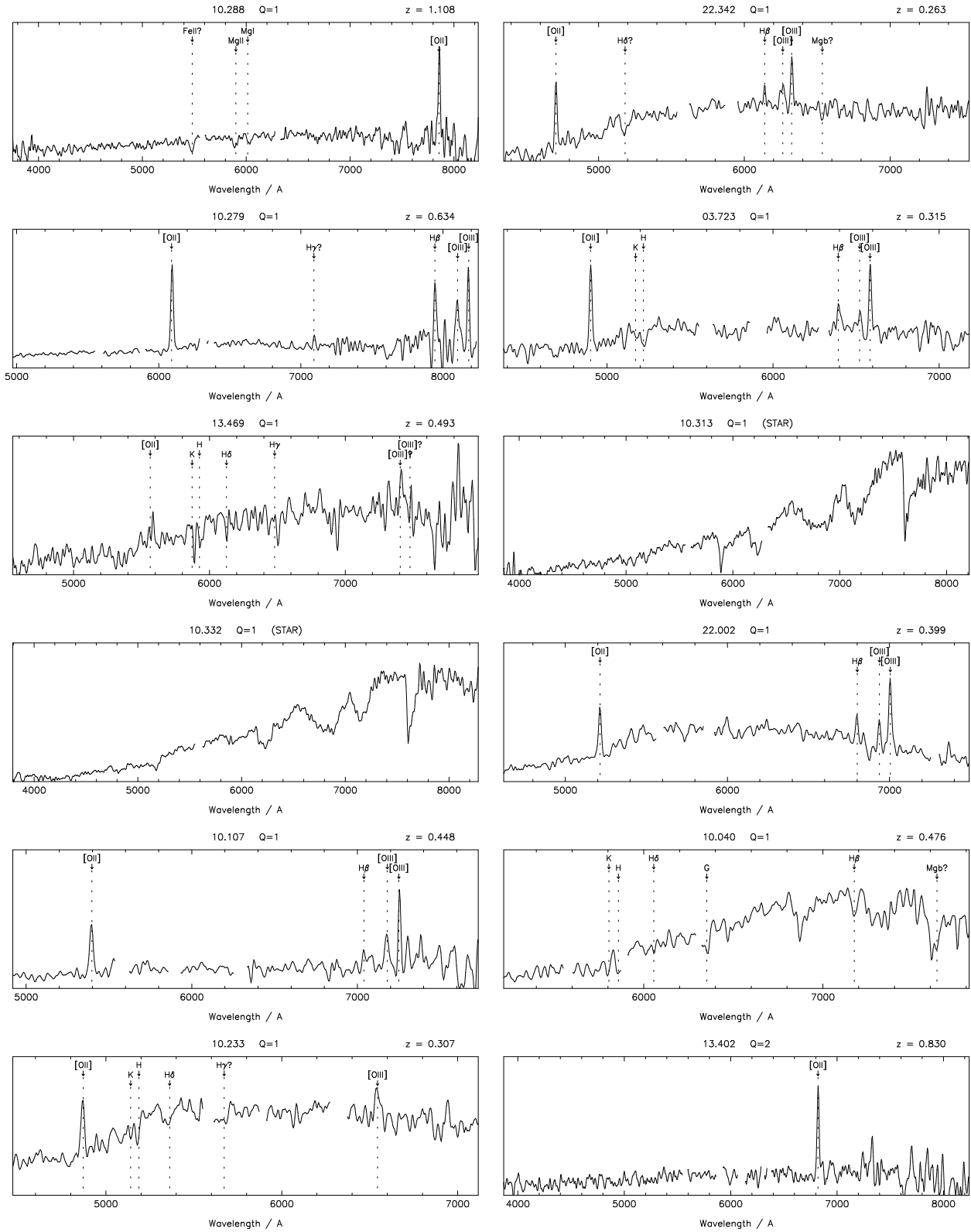

Fig. 1

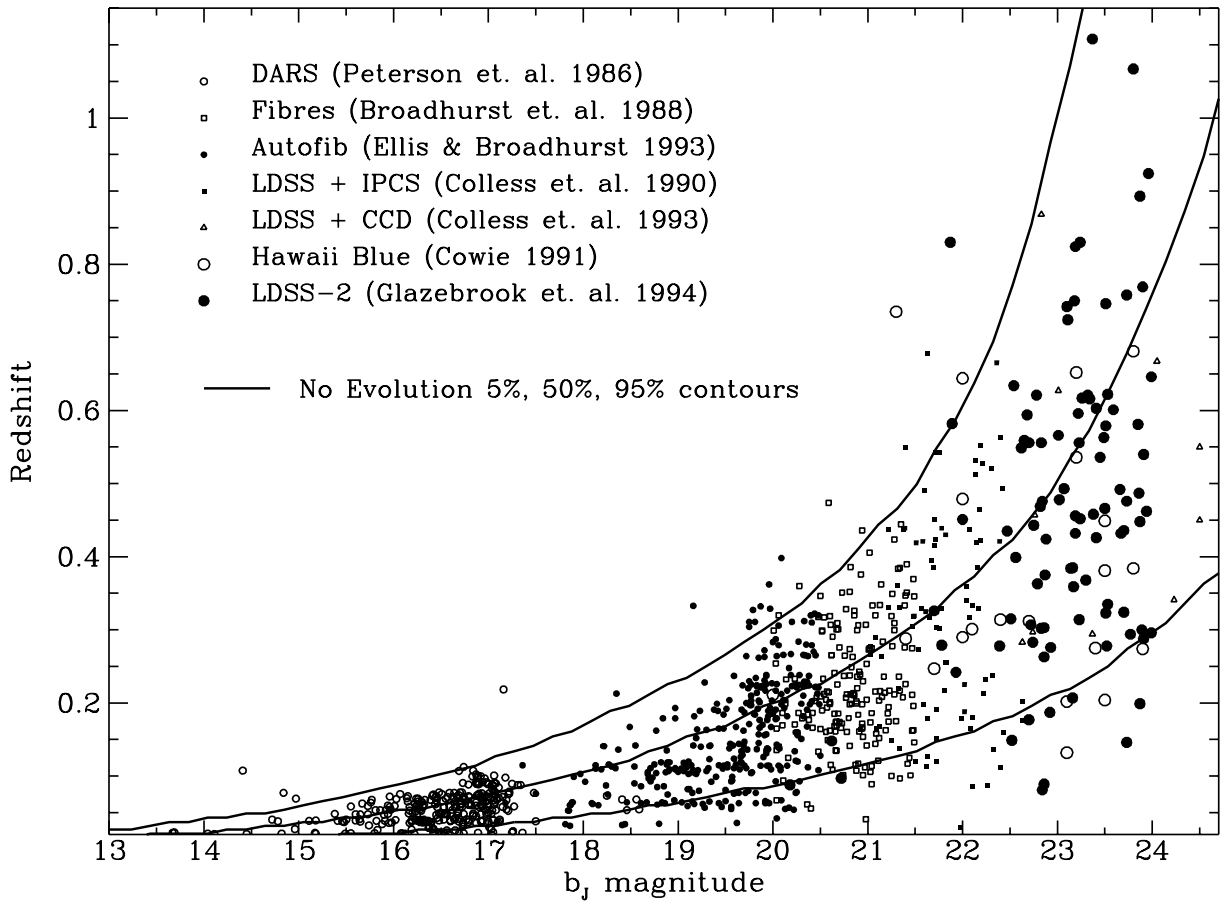

Fig. 2

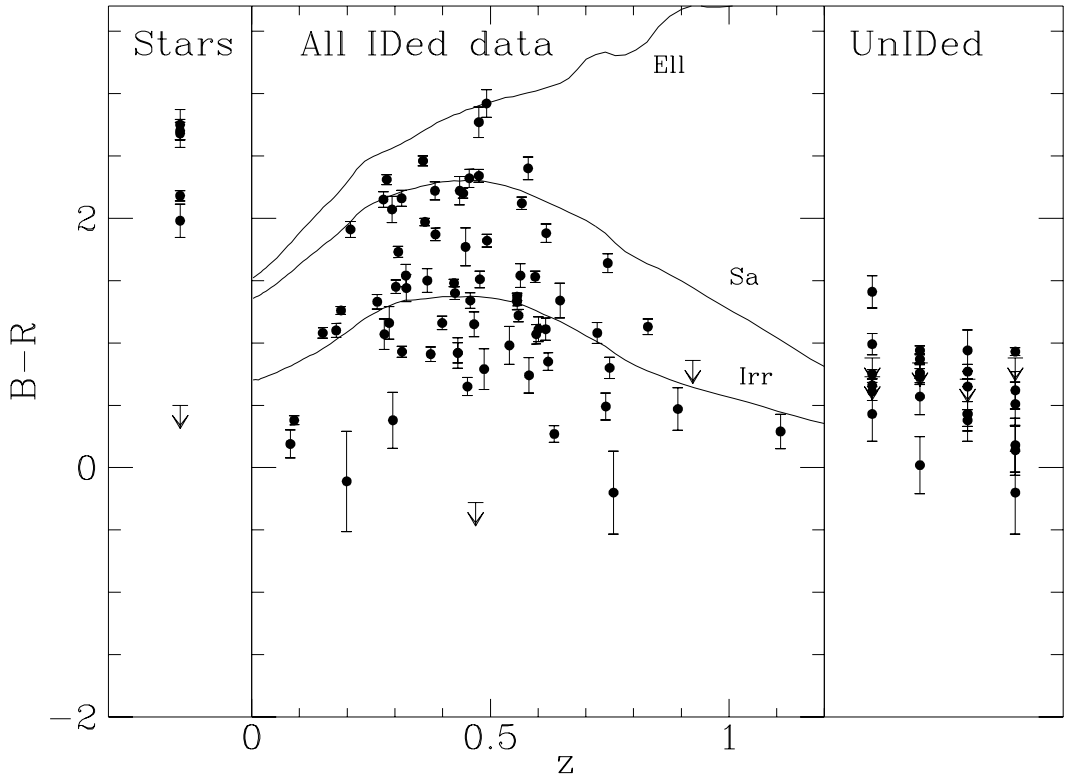

Fig. 3

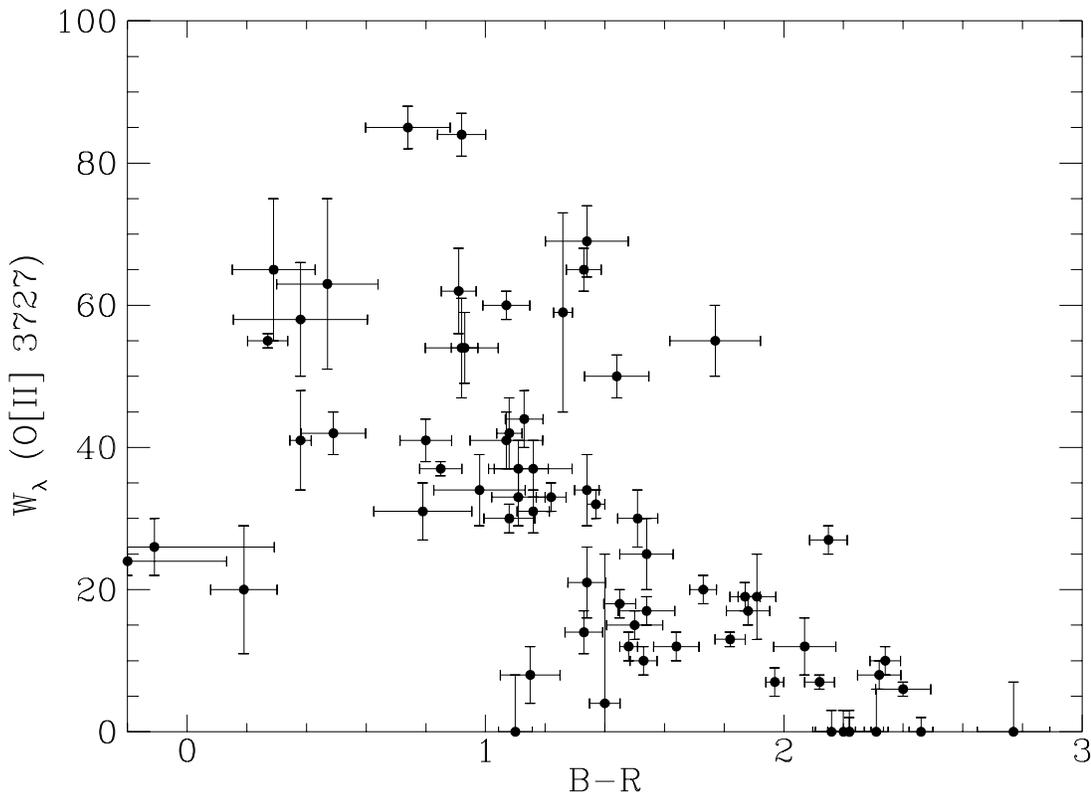

Fig. 4

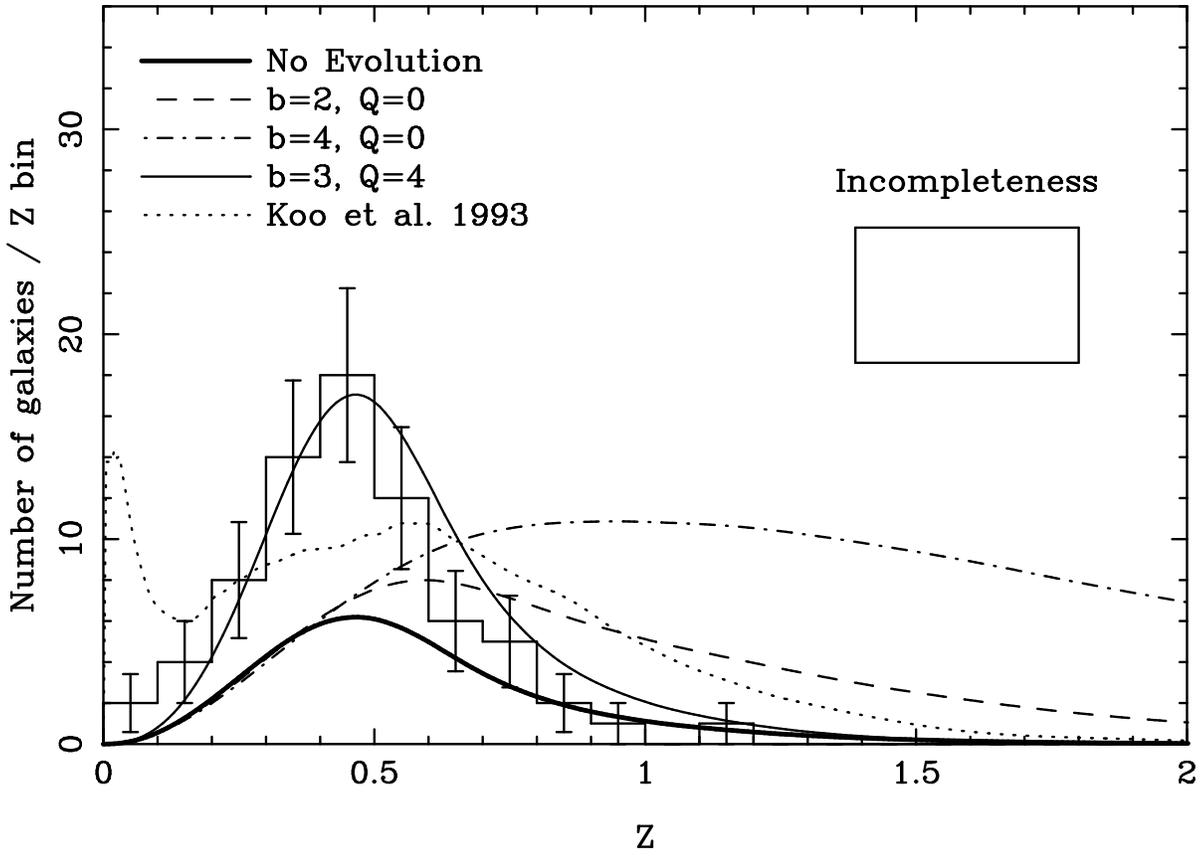

Fig. 5a

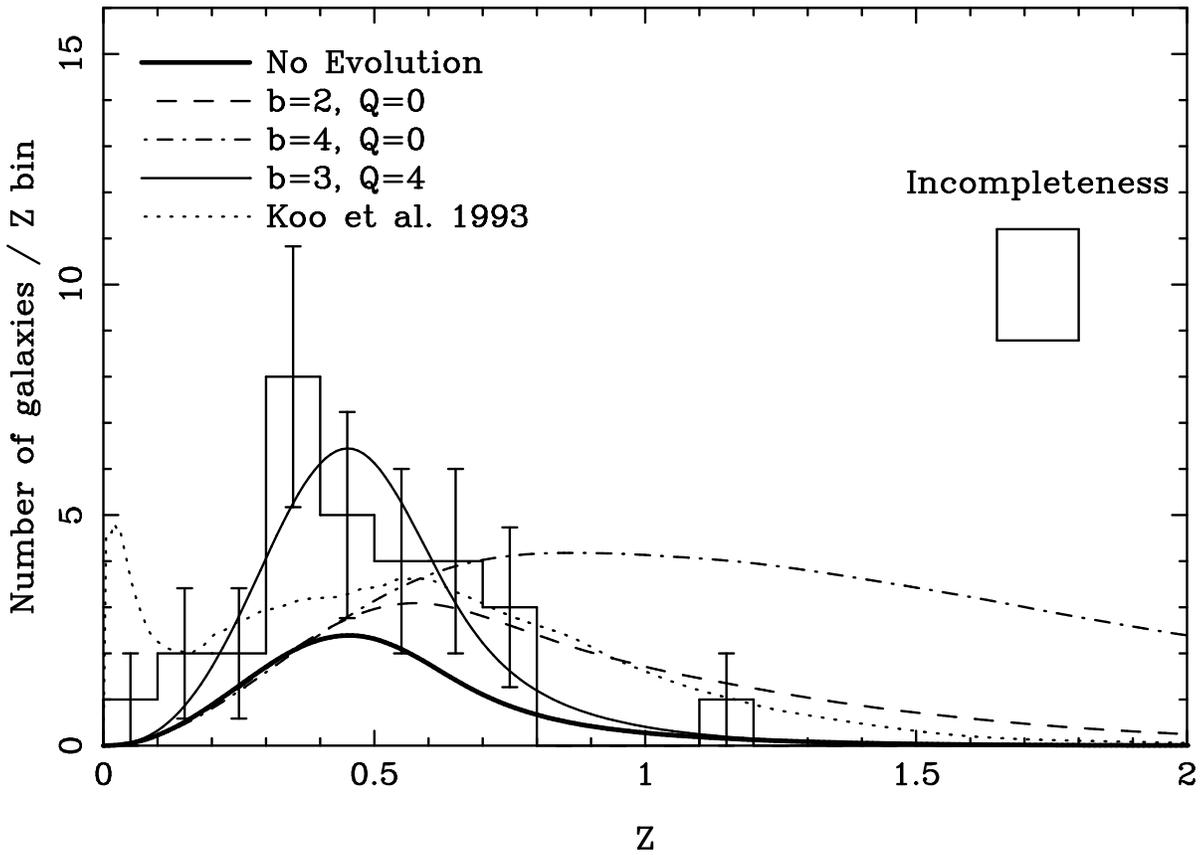

Fig. 5b